\documentclass{amsproc}

\usepackage{amssymb}

\usepackage{graphicx}

\theoremstyle{definition}

\theoremstyle{remark}

\newcommand{\ba}{\begin{align}}

\newcommand{\bea}{\begin{eqnarray}}
\newcommand{\eea}{\end{eqnarray}}
\newcommand{\be}{\begin{equation}}
\newcommand{\ee}{\end{equation}}

\def\Mb{\mathbf{M}}
\def\Sb{\mathbf{S}}
\def\Ab{\mathbf{A}}
\def\Fb{\mathbf{F}}
\def\Db{\mathbf{D}}

\newcommand{\cD}{{\mathcal D }}

\def\Tr{{\rm Tr}}

\newcommand{\ma}{{\textrm{ch}}}
\newcommand{\ml}{\chi}

\def\Deltamax{\Delta_1}
\def\nD{n_{{\cD }}}
\def\bo{\Deltamax}
\def\htot{h^\tot}

\def\tot{\mathrm{tot}}

\newcommand{\ie}{{\it i.e.~}}

\numberwithin{equation}{section}

\begin{document}

\title{Modularity, Calabi-Yau geometry and 2d CFTs}


\author{Christoph A.~Keller}
\address{NHETC and Department of Physics and Astronomy\\
Rutgers, The State University of New Jersey\\
Piscataway, New Jersey 08854-8019, USA.}
\curraddr{}
\email{keller@physics.rutgers.edu}
\thanks{It is a great pleasure to thank H.~Ooguri and D.~Friedan
for collaboration on the work described here, and T.~Hartman for discussions
of some of the points presented here. This work was supported by the Rutgers
New High Energy Theory Center and by U.S. DOE Grants No.~DOE-SC0010008,
DOE-ARRA-SC0003883 and 
DOE-DE-SC0007897.
}

\subjclass[2010]{81T40,11F03}

\date{}

\begin{abstract}
We give a short overview over recent work on finding constraints
on partition functions of 2d CFTs from modular invariance. We summarize
the constraints on the spectrum and their connection to Calabi-Yau
compactifications.
\end{abstract}

\maketitle

\section{The modern bootstrap and modular invariance}
The conformal bootstrap is the project of
constructing conformal field theories from 
consistency conditions imposed by conformal invariance 
\cite{Polyakov:1970xd,Migdal:1972tk,Polyakov:1974gs}.
Compared to other methods, its main advantage is that it does
not rely on a Lagrangian description of the CFT.
It makes use of the conformal symmetry
of the theory by decomposing amplitudes into conformal blocks ---
the contributions of the irreducible representations of the conformal 
group.
In principle it is thus possible to classify and construct
all CFTs, including strongly coupled ones.

For CFTs in two dimensions, the amplitudes
are subject to two types of consistency
conditions. First, there is crossing symmetry. Defining
the four point function
\be
G(x) := \langle\phi(\infty)\phi(1)\phi(x)\phi(0)\rangle\ ,
\ee
it follows from invariance under the global conformal transformation
$z \mapsto 1-z$ that $G(x)=G(1-x)$. On the other hand 
one can decompose $G$ into conformal blocks
\be\label{cross}
G(x) = \sum_{\psi \in S} C^2_{\phi\phi\psi} \mathcal{F}(x)\mathcal{F}(\bar x)\ .
\ee
The conformal block $\mathcal{F}$ is universal and only depends
on the weights $h$ and the central charge $c$. 
Here $S$ is the spectrum of all primary fields
that appear in the OPE of $\phi$ with itself.
Crucially, $\mathcal{F}(x) \neq \mathcal{F}(1-x)$.
This means that (\ref{cross}) can only be satisfied if we choose
the spectrum $S$ and the three-point functions $C_{\phi\phi\psi}$
very carefully. For rational CFTs there are only a finite
number of primaries, so that finding solutions is feasible.
More recently, starting with \cite{Rattazzi:2008pe} 
there has been a lot of progress also for
the general case and for higher dimensions. 
The main ingredients for this modern bootstrap approach to work
are having explicit
expressions for $\mathcal{F}(x)$ and unitarity of the theory, which implies
$C^2 \geq 0$. Fixing the dimension of the external field
$\phi$, this allows to deduce an upper bound for 
the lowest lying field in the spectrum of the OPE of
$\phi$ with itself.

The second consistency condition is modular invariance for amplitudes on
higher genus Riemann surfaces.
In particular it requires that the partition function
\be\label{partfunction}
Z(\tau) = \Tr q^{L_0-c/24} \bar q^{\bar L_0-c/24} \qquad q = e^{2\pi i\tau}
\ee
is invariant under modular transformations, namely $Z(-1/\tau) = Z(\tau)$.
This follows from interpreting (\ref{partfunction}) as
the vacuum amplitude on the torus. Since this should only
depend on the conformal structure $\tau \in \mathbb{H}^+/SL(2,\mathbb{Z})$
of the torus,
it must be a modular invariant function of $\tau$.
Again the representation theory of the conformal group
allows us to decompose $Z$ into characters,
\be
Z(\tau) = \sum_{h,\bar h\in S} N_{h,\bar h}\overline{\chi_{\bar h}(\tau)}\chi_h(\tau)\ ,
\ee
from which we can derive constraints on the spectrum $S$
and the multiplicities $N_{h,\bar h}$.
As before, the characters $\chi$ themselves
are not modular invariant, so that only special
choices for the spectrum give a good partition function.

Note that for 2d CFTs crossing symmetry on the sphere 
and modular invariance on the torus are enough to ensure
consistency of the theory \cite{Moore:1988qv}. 
We can obtain amplitudes on higher genus
Riemann surfaces by gluing these two
components, the consistency of this procedure
being guaranteed by crossing symmetry
of the 4pt functions and modular invariance
of the torus 1pt functions. The situation is less clear
in higher dimensions, and it is still an open
question what corresponds to modular invariance
then.

We will describe here the investigation 
\cite{Hellerman:2009bu,Keller:2012mr,Friedan:2013cba} of CFTs with
a gap. For such theories, other than the vacuum we do not allow
for any primary with total weight less than a certain
weight $\Delta_1$,
so that $S$ is of the form
\be
S_{\Delta_{1}} = \left\{ (h,\bar h) \, : \,
h,\bar h \ge 0 , \;
h + \bar h \ge \Delta_{1}\right\}\ .
\label{eq:SDelta1}
\ee
A theory which saturates the largest possible gap
is called extremal.
Such extremal theories could for instance arise as a holographic dual to
pure gravity on $AdS_3$ \cite{Witten:2007kt}. More generally, this gives
information on the structure of modular forms in the following
sense.

Consider first the analog problem for meromorphic partition
functions. Modular invariance means that $Z(\tau)$ is well-defined
on $\mathbb{H}^+/SL(2,\mathbb{Z})$, which is compact. 
Meromorphic functions on compact spaces are determined
by their poles. Since for physical reasons we know that the
only pole is at $q=0$, it follows that for a holomorphic
CFT with central charge $c=24k$, the largest gap
is $\Delta_1 = \frac{c}{24}+1$. Note that although
one can construct such extremal partition functions for all values
of $k$, it is not clear that the corresponding extremal
CFTs exist.

We expect the space of non-meromorphic partition functions
to be much bigger and its structure more complicated.
A priori the bound on the gap will thus be weaker,
and it is interesting to investigate by how much.
In the first part of this review we will discuss this
question.

As an aside, note that for such holomorphic theories 
also the crossing symmetry problem
simplifies. In this
case the 4pt function is meromorphic and thus determined
by its poles. For an external field of weight $h_\phi$
the first $2h_\phi-1$ terms in the OPE thus fix the
full 4pt function. The space of contributions to 
the 4pt function is thus at most $2h_\phi-1$ dimensional,
and it is a problem in finite dimensional linear
algebra to find crossing symmetric elements.
In the end one finds an upper bound for the 
field in the internal channel of the form
$h_\psi \leq \frac{4}{3}h_\phi + O(1)$,
as was obtained in \cite{Bouwknegt:1988sv}
in the context of classifying $W$-algebras.
The advantage of this method is that there
are no numerical computations involved,
which in particular allows to go to very high
values of $c$ and $h_\phi$.
It would be interesting to extend this type of 
analytic approach to non-holomorphic theories.

Finally note that so far all these methods only
give an upper bound for the lowest field in the
spectrum. Once we are confident that we have
obtained the best possible bound, we can try to
reconstruct the spectrum of this extremal theory.
This is done by extracting the
multiplicity of the lowest primary fields,
and then bootstrapping our way up the spectrum.
In various contexts this type of program has been applied
successfully \cite{ElShowk:2012hu,Friedan:2013bha,Qualls:2013eha}.
For our type of problem however generically this approach fails.
As we will see, we do not impose integrality of
the multiplicities $N_{h,\bar h}$ when deriving
our bounds. This means that in general the multplicities
obtained for our extremal spectra will not be integral.
Still in some special cases they may become integer,
which would suggest strongly that there exists
a corresponding CFT.

\section{CFTs with large gaps}
In this section we discuss how to obtain an upper bound 
$\bo$ for the lowest lying primary of a general CFT
of central charge $c$. A first such bound was
obtained in \cite{Hellerman:2009bu}, and
\cite{Friedan:2013cba} discussed how to improve
on it systematically.

For simplicity choose a purely imaginary $\tau = i\beta$.
Introducing the notation $\tilde Z = Z(\beta^{-1})$,
modular invariance of the partition function can be written
as
\be \label{main}
v_0:=-(Z_0 - \tilde Z_0) = \sum_{h,\bar h \in S} N_{h,\bar h}(Z_{h,\bar h} - \tilde Z_{h,\bar h})\ ,
\ee
where the left hand side is the contribution of the vacuum,
and the right hand side the contribution of all the other 
primaries.

There is a very nice geometric interpretation of 
(\ref{main}). Restricting to theories that 
have only fields of total weight $h+\bar h \geq \Delta_1$
in their spectrum $S$, the right hand side
forms a convex cone $C_{\Delta_1}$ in the space
of functions in $\beta$. Checking the existence of such a 
partition function 
reduces to checking if the vacuum contribution $v_0$
is in $C_{\Delta_1}$.

\begin{figure}[htbp]
\begin{center}
\includegraphics[height=.3\textwidth]{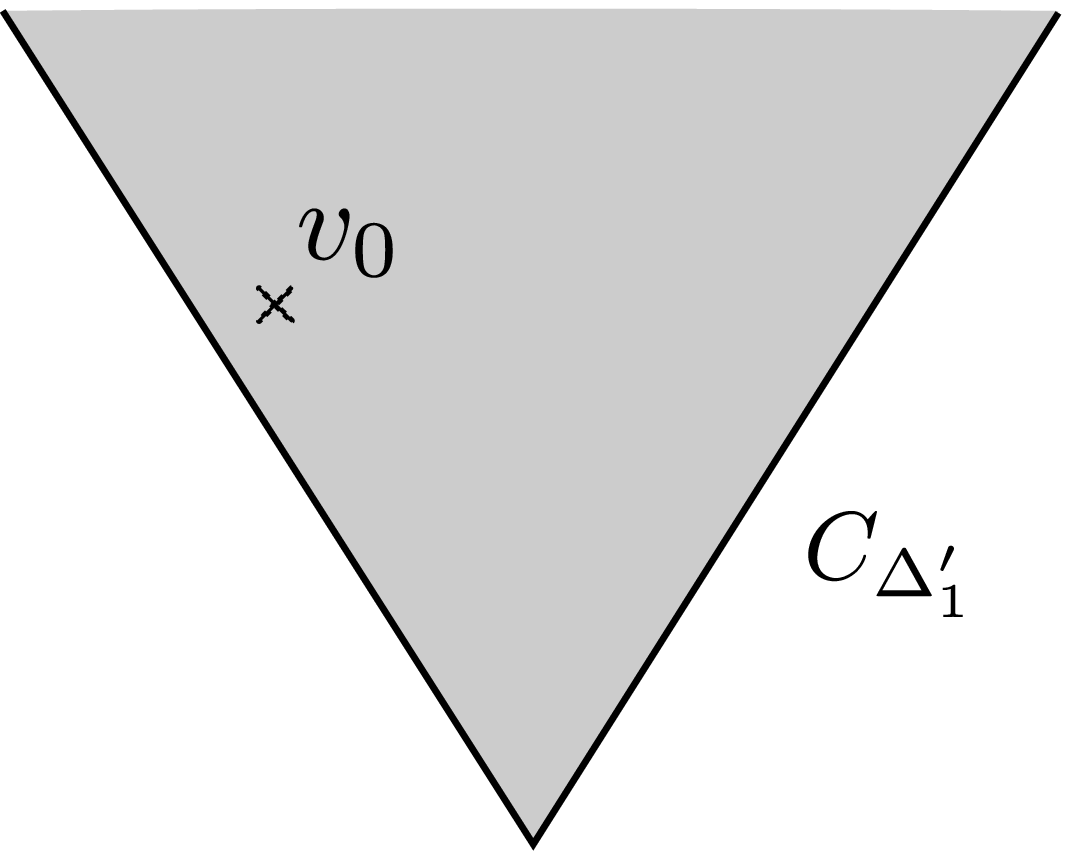}
\includegraphics[height=.3\textwidth]{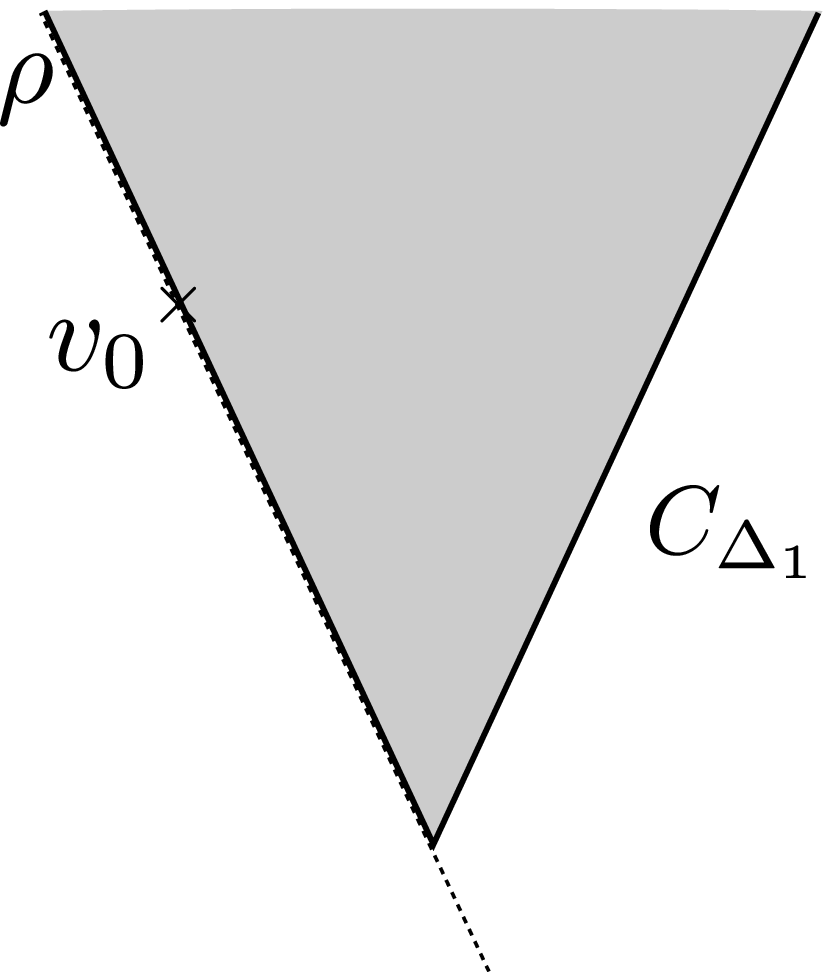}
\includegraphics[height=.3\textwidth]{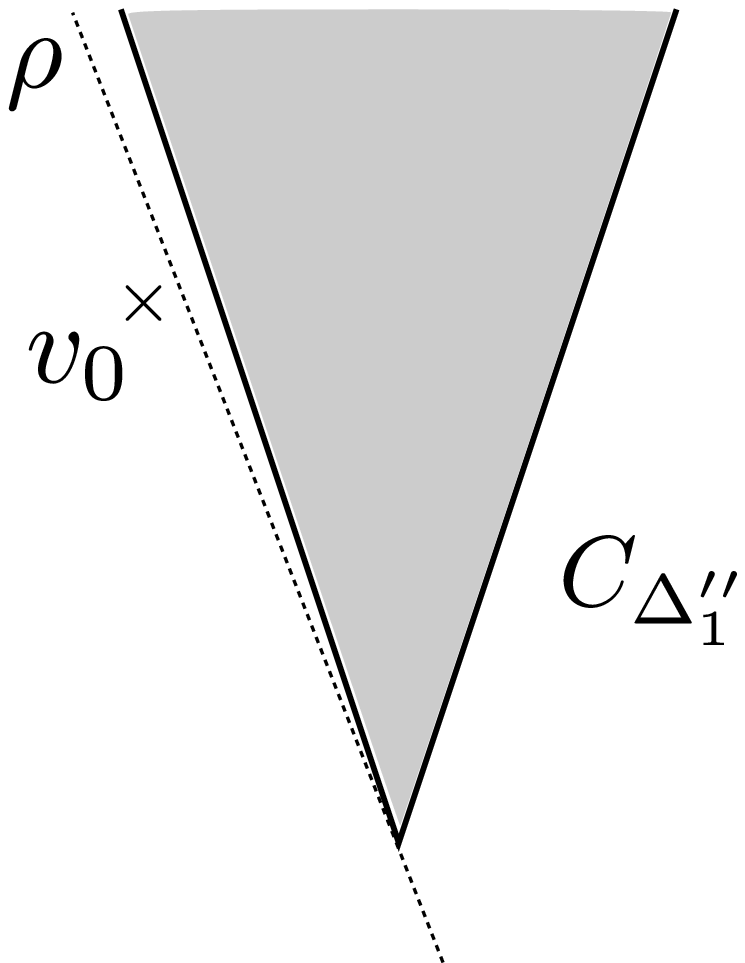}
\caption{The situation for $\Delta_1' < \Delta_1 < \Delta''_1$: For 
$\Delta_1'$, $v_{0}$
is within the cone $C_{\Delta'_1}$ so it is impossible to find a
separating plane. For $\Delta_1''$, $v_{0}$ is outside of the cone, and we
can find a separating plane $\rho$.
}
\label{Fig:cone3}
\end{center}
\end{figure}

One way to check this is to search for a hyperplane
$\rho$ that separates $v_0$ from the cone. Such
a linear functional has to satisfy
\be \label{seprho}
\rho |_{C_{\Delta_1}} \geq 0 \quad \textrm{and}\quad \rho(v_0) < 0\ .
\ee
If we can construct such a functional, then we know that
there cannot be a CFT with a gap as large as $\Delta_1$.
Our strategy will therefore be to scan as systematically
as possible over the space of linear functionals, attempting 
to find a separating one.

To do this, let us first simplify the problem by
defining a reduced partition function
\be
\hat Z(\tau) =|\tau|^{1/2}|\eta(\tau)|^2 Z(\tau)\ .
\ee
Note that $\hat Z$ is still invariant under $S$,
so that we can perform our analysis with the reduced
function. We are using the fact that the characters
of the Virasoro algebra are essentially the $\eta$
function, which itself is modular. The reduced
contribution of the primary fields is the simple monomial
$\hat Z_{h,\bar h} = |\tau|^{1/2}q^{h-\gamma}\bar q^{\bar h-\gamma}$, 
and the vacuum contribution is 
$\hat Z_0 = |\tau|^{1/2}|1-q|^2q^{-\gamma}\bar q^{-\gamma}$.
We have introduced $\gamma=\frac{c-1}{24}$, where
the -1 comes from the effective central charge of the Virasoro
algebra, which we effectively removed by dividing out the $\eta$
function.

Since we are dealing with analytic functions,
one way to construct linear functionals
is by extracting Taylor coefficients.
More precisely we can act with differential
operators $\beta\partial_\beta$ and evaluate
the result at the self-dual point $\beta=1$.
The advantage of using the logarithmic derivative
is that it is odd under $S$.
Given such a differential operator $\mathcal{D}$
of order $n_\mathcal{D}$, by the simple form
of $Z_{h,\bar h}$ we obtain a polynomial $p$,
\be
(\mathcal{D},Z_{h,\bar h}) = e^{-2\pi(\Delta-2\gamma)}p(\Delta)\ .
\ee
The positivity condition in (\ref{seprho}) thus
reduces to checking that $p(\Delta) \geq 0$ if
$\Delta \geq \Delta_1$.
Note that by construction both sides
of (\ref{main}) are odd under $S$. This means
we can restrict to odd differential operators
$\cD$.

We can in fact rewrite any such polynomial
in the form
\be
p(\Delta) = \vec{\Delta}^T Y_1 \vec{\Delta} +(\Delta-\Delta_1)\vec{\Delta}^T Y_2\vec{\Delta}\ ,
\ee
where $\vec{\Delta}=(1,\Delta,\Delta^2,\ldots,\Delta^n)$
and $Y_{1,2}$ are positive semidefinite matrices.\footnote{
In the case here this is related to the fact that
we can write any positive polynomial as a sum of squares.
For polynomials in more variables the situation is more complicated,
and is related to Hilbert's 17th problem.}
To find a separating hyperplane as in (\ref{seprho}),
we thus scan over the space of semidefinite matrices $Y_{1,2}$.
We need to impose some linear constraints ensuring
that the corresponding polynomials actually can come
from a differential operator $\mathcal{D}$, and also
to fix the overall normalization of $\rho$. The goal is
then to find matrices $Y_{1,2}$ which maximize $\rho(-v_0)$.
If this maximum is positive, then we have indeed found
a separating hyperplane and can conclude that $\Delta_1$
is an upper bound for the gap of any CFT.
We have thus phrased our problem as 
maximizing a linear objective function over a set of semidefinite matrices
under a set of linear constraints.
Such problems have been well studied in linear programming, and 
powerful numerical solvers have been developed such as SDPA~\cite{SDPA}. 

\begin{figure}[htbp]
\begin{center}
\includegraphics{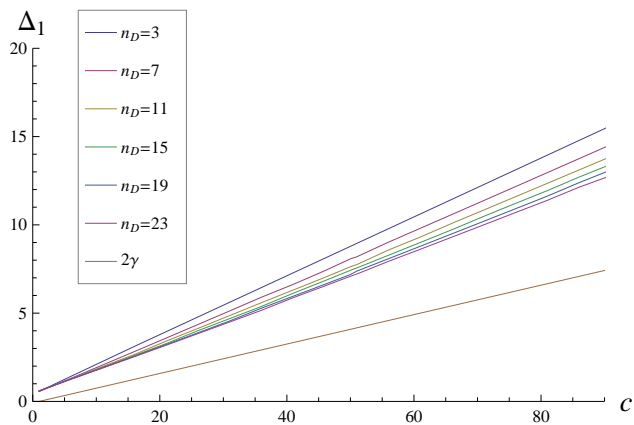}
\caption{The bound $\Delta_1$ as a function of $c$. The bottom line is 
the barrier $2\gamma$,
beyond which one can never improve any bound
from our methods.}
\label{Fig:bosonLinear}
\end{center}
\end{figure}

We have plotted the results in figure~\ref{Fig:bosonLinear},
which shows how the bound
$\Delta_1(c)$ improves as we increase the order
of the differential operator $n_{\mathcal{D}}$.
The original bound \cite{Hellerman:2009bu} is
the case $\nD=3$.
For small $c$ the improvements are relatively small,
and we seem to converge to a best possible bound
very quickly. For larger $c$ we have to go to higher
and higher order operators to get accurate results.

Let us discuss the asymptotic behavior for large $c$ 
in more detail here.
The main interest for this comes from the $AdS_3/CFT_2$
correspondence, in which the limit $c\rightarrow\infty$
on the CFT side corresponds to the classical limit
on the gravity side, for which semiclassical computations
are appropriate. The extremal partition function we investigate
here would correspond to pure gravity \cite{Witten:2007kt}.
More generally, theories that are dual to gravity theories
should have only a small number of low-lying states.
It is thus reasonable to assume that to leading order
they have to satisfy the same bound $\Delta_1$.
On the other hand all such theories
should contain BTZ black holes \cite{Banados:1992wn}.
On the CFT side this means that there should be
an exponential number of primary fields starting
at weight $\frac{c}{12}$. This suggests that 
$\Delta_1$ should grow as $\frac{c}{12}$.

The original bound found for $\nD=3$
grows as $\frac{c}{6}$. Figure~\ref{Fig:bosonLinear}
seems to suggest that the bounds from higher
operators have a better asymptotic. It turns out however 
that for large enough $c$ they all asymptote as
$\frac{c}{6}$ if we keep $\nD$
fixed. Since figure~\ref{Fig:bosonLinear} 
is so suggestive, it is possible that this simply an order
of limit issue. That is, for a given $c$ it may
be possible to find a differential operator that gives
a bound whose leading behavior in $c$ is better
than $\frac{c}{12}$ --- our arguments simply show
that the order of this operator would have
to grow with $c$. This suggests that as
a basis for linear functionals, finite order
differential operators are not really
appropriate for investigating this type of question.
It would be interesting to construct a more
appropriate basis, and check if the asymptotic
behavior is indeed as predicted by holography.

The only thing we do know for certain is that no 
bound obtained in such a way
can be better than the barrier $\Delta_1 = 2\gamma = \frac{c}{12}+O(1)$. This can
be seen for example by comparison to the holomorphic case
discussed in section~1,\footnote{To compare
to the holomorphic case, note that in
the results quoted here we set $c=\frac{c_L+c_R}{2}$.}
or also by a direct argument \cite{Friedan:2012jk}.

\section{Calabi-Yau compactifications}
Let us now consider non-linear $\sigma$-models
on Calabi-Yau $d$-folds. Such CFTs have
more symmetry, namely extended $N=(2,2)$
superconformal symmetry. The extension
comes from the holomorphic $(d,0)$
form $\Omega$, which on the CFT side
leads to invariance under one unit of
spectral flow. Because of the $U(1)$ 
$R$-symmetry of the algebra, the Cartan
torus has an additional element, so that we can
compute a generalized partition function
\be
Z(\tau,\bar\tau,z,\bar z) = \Tr \left (  q^{L_0-c/24}\bar q^{\bar L_0-c/24}
y^{J_0} \bar y^{ \bar J_0}\right )\ ,\qquad y=e^{2\pi i z}\ .
\ee
We can compute the partition function for the different
spin structure by taking the trace in the NS or R sector,
and by inserting fermion parity operators.
We will concentrate on the $(-,-)$ spin structure, 
that is the untwisted NS sector without any fermion
number operator inserted.
Under the $S$ transformation 
\be
S: (\tau, z) \mapsto (\tilde \tau,\tilde z)= (-1/\tau,z/\tau)\,,
\ee
it transforms as \cite{Kawai:1993jk}
\be \label{charTrafo}
Z(\tau,z) =  e^{-2\pi i \frac{d}{2} \frac{z^2}{\tau}} \,
e^{2\pi i \frac{d}{2} \frac{\bar z^2}{\bar \tau}}
\,Z(\tilde \tau,\tilde  z)\, ,
\ee
\ie it is invariant up to a phase.
Note that it will not transform to itself under the full modular group,
but rather under $\Gamma_\theta$, the subgroup generated by
$S$ and $T^2$. Using a combination of $T$ and $S$ transformations
we can obtain partition functions of other spin structures.

The representation theory of the extended
$N=(2,2)$ superconformal symmetry has
been studied in \cite{Odake:1988bh}. There
are $d$ short representations $\ml^Q$,
and $d-1$ families of long representations
$\ma^Q_h$, labeled by their weights $h$
and $U(1)$ charges $Q$. Combining left-
and right-moving short representations gives
$\frac{1}{2}$-BPS states, whereas a combination
of a long with a short representation give a
$\frac{1}{4}$-BPS state, leading to a
total partition function of the form
\be
Z = Z_{\frac{1}{2}BPS}+Z_{\frac{1}{4}BPS}+Z_m\ ,
\ee
where $Z_m$ contains all the non-BPS states.
Schematically, the representations of the $\sigma$-model 
are related to the Calabi-Yau as 
\begin{eqnarray*}
\textrm{topology} &\leftrightarrow& \textrm{BPS states}\\
\textrm{geometry} &\leftrightarrow& \textrm{non-BPS states}
\end{eqnarray*}
More precisely, the $\frac{1}{2}$BPS states are
fixed by the Hodge numbers, and the $\frac{1}{4}$BPS states
essentially by the elliptic genus of the manifold.
In general a lot is known about the topology
of CY manifolds, and hence about the BPS states
of the theory. We will try to extract information
about the spectrum of non-BPS states. In the end,
we will give an upper bound $\bo$ for the lowest
lying non-BPS primary as a function of the Hodge
numbers of the manifold.

For this analysis it is straightforward to apply the
same method as in the Virasoro case.
A priori it seems that the space of
functions to analyze now depends on
the two variables $\tau$ and $z$. However,
spectral flow invariance allows us to
factor out the $z$ dependence of $Z$ in
terms of $\theta$-functions. 
More precisely, Hermite's Lemma tells us that the 
space of such functions
has dimension $d$ over the functions of $\tau$,
and can be spanned by $d$ $\theta$-functions
\be
f_{d}^{Q}(\tau,z)= 
\frac{1}{\eta(\tau)}\sum_{m\in\mathbb{Z}}q^{\frac{d}{2}(m+Q/d)^2}y^{d(m+Q/d)}
\ .\ee
The $f_{d}^{Q}$ we have picked here form a particular suitable 
basis because they transform
linearly under the $S$ modular transformation with
simple transformation matrices:
Let $\Fb_{d}$ be the $d$-vector with entries $e^{\frac{i\pi dz^2}{2\tau}}f_{d}^{Q}$.
Then \cite{Odake:1989ev}
\be\label{ftrafo}
\Fb_{d} (\tilde \tau,\tilde z) = \Sb _{d}\,\Fb_{d} (\tau,z)
\,,\qquad
(\Sb _{d})_{Q'}^Q = d^{-\frac12} e^{-2\pi i QQ'/d}\, .
\ee
The transformation matrix $\Sb_d$ is unitary and
its entries are just numbers.
In view of (\ref{charTrafo}), we again define a reduced
partition function
\be
\hat Z(\tau,z) = \left| e^{\frac{i\pi dz^2}{2\tau}}
(-i\tau)^{1/4}\eta(\tau)\right|^2 Z(\tau,z)\, ,
\label{eq:Neq2Zreduced}
\ee
so that $S$ modular invariance becomes simply
\be
 \hat Z(\tau,z) = \hat Z(\tilde \tau,\tilde z) \,.
\label{eq:Neq2Zreducedmodinv}
\ee
Using the fact that we can express all
characters in terms of $\Fb$, 
we can write the reduced partition function as
\be\label{ZCYred}
\hat Z(\tau,z) = \Fb_{d}^\dagger \,\Mb (\tau)\,\Fb_{d}
\,,
\ee
where the $d\times d$ matrix $\Mb(\tau) $ is determined by the 
multiplicities of the various representations. 
Note that crucially $\Mb(\tau)$ 
only depends on $\tau$, and not on $z$ anymore.
In some cases the expressions for $\Mb$ become very simple,
and we will give explicit examples below.
Modular invariance (\ref{eq:Neq2Zreducedmodinv}) 
is then equivalent to the matrix equation
\be \label{matrixeq}
\Mb(\tau) = \Sb _{d}^\dagger\, \Mb(\tilde \tau)\, \Sb _{d}\,.
\ee
We are thus back at a variant of the bosonic modular invariance 
problem. Linear functionals $\rho$ are now given
by $d\times d$ matrices 
$\cD^{QQ'}$ of differential operators 
\be
\cD = \Db(\tau\partial_{\tau})
\ee
where $\Db$ is a matrix of polynomials in $\tau\partial_{\tau}$ and 
$\bar \tau\partial_{\bar \tau}$.
A matrix differential operator $\cD$ acts on a matrix of functions 
$\Ab$ by
\be
(\cD,\Ab) = \Tr (\Db^\dagger\Ab)\bigr|_{\tau=i}\,.
\ee
We separate $\Mb$ into two parts
\be
\Mb = \Mb_{BPS} + \Mb_{m}\, ,
\ee
where the BPS contribution $\Mb_{BPS}$ comes from the multiplicities 
that are determined by the known topological properties of the 
Calabi-Yau manifold, that is the BPS states.  
The rest of the multiplicities determine 
$\Mb_{m}$, which due to their positivity again
span a convex cone in the function space.  
On this cone the semidefinite condition on $\cD$ is
\be
(\cD,\Mb_{m})\ge 0 
\label{eq:semidefinitegenerald}
\ee
for all possible multiplicities consistent with a given gap
$\Delta_{1}$.

We shall restrict ourselves to the case of Calabi-Yau 3-folds
here. In this case additional simplifications occur.
The $\frac{1}{2}$BPS part is then given by\footnote{For simplicity
we assume here that there are no additional continuous symmetries in the theory.}
\be
\label{eq:ZhalfBPS}
Z_{\frac{1}{2}BPS} = \bar{\ml}^0\ml^0+ h^{1,1}( \bar{\ml}^{1}\ml^1 + \bar{\ml}^{-1}\ml^{-1})
+ h^{2,1} (  \bar{\ml}^{1}\ml^{-1}+\bar{\ml}^{-1}\ml^1) \ .
\ee
The elliptic genus is given by
\be
J=(h^{1,1}-h^{2,1})\phi_{0,3/2}
\ee
where $\phi_{0,3/2}$ is the unique weak Jacobi form of
weight 0 and index $3/2$.
Rather surprisingly it turns out that in the case at hand
(\ref{eq:ZhalfBPS}) itself is already a weak Jacobi form once we flow to the
R sector and compute the elliptic genus, \ie
\be
Z_{\frac{1}{2}BPS} \rightarrow J\ .
\ee
It follows that the $\frac{1}{4}$BPS
contributions need to vanish, so that $Z_{\frac{1}{4}BPS}=0$. 
Note that for other dimensions
$d$ one has add corrections from $\frac{1}{4}$BPS 
to obtain a weak Jacobi form. It would be interesting to
understand this coincidence for $d=3$ more conceptually.

This allows us to reduce the matrix $\Mb$ from a 
$3\times3$ matrix to a $2\times2$ matrix. The 
massive contribution $\Mb_m$ in particular
has the very simple form
\be
\Mb_{r}(\tau)=   |\tau|^{1/2}\sum_{ h,\bar h}\left(\begin{array}{cc}
(N^{m}_{\bar h h})_{00}\,\bar q^{\bar h-1/4}q^{h-1/4}
&(N^{m}_{\bar h h})_{01}\,\bar q^{\bar h-1/4 } q^{h-1/2}\\[1ex]
(N^{m}_{\bar h h})_{10}\,\bar q^{\bar h-1/2}q^{h-1/4}
&(N^{m}_{\bar h h})_{11}\,\bar q^{\bar h-1/2}q^{h-1/2}
\end{array}\right)\,. \label{Mm}
\ee
As all the entries are simple monomials, we are back
to the bosonic case.

\begin{figure}[htbp]
\begin{center}
\includegraphics{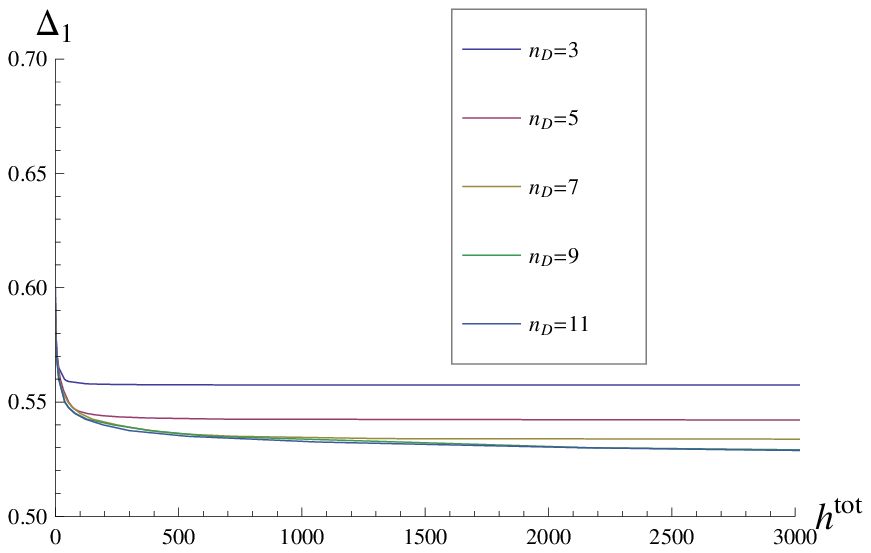}
\caption{$\bo(\htot)$ for various $\nD$.}
\label{Fig:N2diffops}
\end{center}
\end{figure}

In figure~\ref{Fig:N2diffops} we have plotted
the bound for various values of the order $\nD$
of the differential operator. 
As expected, 
$\bo$ is monotonically decreasing in $\htot$.
We find that $\bo$
converges very quickly in $\nD$ for small Hodge numbers.
The weakest bound is for vanishing Hodge numbers,
for which we find $\bo < 0.60$. Note in particular
that this means that the lowest lying state is always
a non-BPS state.

For large Hodge numbers we have to go to higher
and higher order to get a reliable result.
This suggests that it is more efficient
to perform a Taylor expansion around a point
other than $\beta=1$. Using this one can in fact
show that $\bo \rightarrow \frac{1}{2}$ for
$\htot \rightarrow \infty$. Note that 
$\frac{1}{2} = 2\frac{c-c_{eff}}{24}$ corresponds
to the $2\gamma$ we encountered in the bosonic case,
where now $c=9$ and the effective central charge
of the $N=2$ algebra is $c_{eff}=3$.

Let us briefly discuss the interpretation of this result.
The bound we have found is uniform over the whole
moduli space of a Calabi-Yau. As a toy model
for this type of bound,
take a free theory compactified on the torus.
For large radius, the bound is satisfied by
KK momentum modes. For very small radius, 
it is satisfied by the winding modes. Over the
whole moduli space the gap becomes biggest
at the self-dual point. Similarly, in general
our bound $\Delta_1$ will be easily satsified
by the eigenmodes of the Laplacian
in the large volume limit. For small volumes,
presumably it will again be easily satisfied
by whatever corresponds to winding modes.
We expect our bound to be most constraining
in the highly stringy regime. One can
conjecture the existence of a generalized
self-dual point where our bound is actually saturated.

Note also that in principle this result makes
a phenomenological prediction for type II
compactifications, namely the existence of
at least one stringy state below a certain
threshold. This is a very weak bound since
by the remarks above this state will have
mass of the order of the string scale. Nonetheless
this is one of very few $\alpha'$ exact results
that we are aware of. For a proper treatment
one would have to perform the GSO projection,
which will eliminate the tachyonic states
with $h+\bar h <1$ that seem to appear in
the spectrum so far.

Ultimately we would like to be able to rule
out theories completely. One way we could have
achieved that if the bound $\Delta_1$ had become
negative for large enough $\htot$, which would
have put an upper bound on the Hodge numbers
of a CY 3-fold. This would give strong evidence
in favor of the conjecture that there are
only finitely many topological families
of such manifolds \cite{Yau:2000}.

Even though there is still some
room for improvement on the bounds obtained
from the partition function, it seems unlikely
that this would be strong enough. We can
show however that the number of states
below $\bo$ grows linearly in $\htot$
so that the spectrum becomes continuous as
$\htot\rightarrow\infty$. We usually associate
such a behavior with a decompactification
limit, and by combining this result with 
other methods it may be possible to show inconsistency.

Ultimately what one should do is to probe
deeper by combining modular invariance
on the torus in an effective way with crossing
symmetry on the sphere. This will presumably
also allow to include other topological information
such as the chiral ring of the CFT.


\bibliographystyle{amsplain}

\bibliography{../ref}

\providecommand{\bysame}{\leavevmode\hbox to3em{\hrulefill}\thinspace}
\providecommand{\MR}{\relax\ifhmode\unskip\space\fi MR }
\providecommand{\MRhref}[2]{%
  \href{http://www.ams.org/mathscinet-getitem?mr=#1}{#2}
}
\providecommand{\href}[2]{#2}
\begin{thebibliography}{10}

\bibitem{Banados:1992wn}
Maximo Banados, Claudio Teitelboim, and Jorge Zanelli, \emph{{The Black hole in
  three-dimensional space-time}}, Phys.Rev.Lett. \textbf{69} (1992),
  1849--1851.

\bibitem{Bouwknegt:1988sv}
Peter Bouwknegt, \emph{{EXTENDED CONFORMAL ALGEBRAS}}, Phys.Lett. \textbf{B207}
  (1988), 295.

\bibitem{ElShowk:2012hu}
Sheer El-Showk and Miguel~F. Paulos, \emph{{Bootstrapping Conformal Field
  Theories with the Extremal Functional Method}},  (2012).

\bibitem{Friedan:2013cba}
Daniel Friedan and Christoph~A. Keller, \emph{{Constraints on 2d CFT partition
  functions}}, JHEP \textbf{1310} (2013), 180.

\bibitem{Friedan:2012jk}
Daniel Friedan, Anatoly Konechny, and Cornelius Schmidt-Colinet, \emph{{Lower
  bound on the entropy of boundaries and junctions in 1+1d quantum critical
  systems}}, Phys.Rev.Lett. \textbf{109} (2012), 140401.

\bibitem{Friedan:2013bha}
\bysame, \emph{{Precise lower bound on Monster brane boundary entropy}}, JHEP
  \textbf{1307} (2013), 099.

\bibitem{Hellerman:2009bu}
Simeon Hellerman, \emph{{A Universal Inequality for CFT and Quantum Gravity}},
  JHEP \textbf{1108} (2011), 130.

\bibitem{Kawai:1993jk}
Toshiya Kawai, Yasuhiko Yamada, and Sung-Kil Yang, \emph{{Elliptic genera and
  N=2 superconformal field theory}}, Nucl.Phys. \textbf{B414} (1994), 191--212.

\bibitem{Keller:2012mr}
Christoph~A. Keller and Hirosi Ooguri, \emph{{Modular Constraints on Calabi-Yau
  Compactifications}}, Commun.Math.Phys. \textbf{324} (2013), 107--127.

\bibitem{SDPA}
M.~Fukuda K.~Nakata M.~Yamashita, K.~Fujisawa and M.~Nakata, \emph{{A
  high-performance software package for semidefinite programs: SDPA 7}},
  Research Report B-463, Dept. of Mathematical and Computing Science, Tokyo
  Institute of Technology, Tokyo, Japan, September 2010.

\bibitem{Migdal:1972tk}
Alexander~A. Migdal, \emph{{Conformal invariance and bootstrap}}, Phys.Lett.
  \textbf{B37} (1971), 386--388.

\bibitem{Moore:1988qv}
Gregory~W. Moore and Nathan Seiberg, \emph{{Classical and Quantum Conformal
  Field Theory}}, Commun.Math.Phys. \textbf{123} (1989), 177.

\bibitem{Odake:1988bh}
Satoru Odake, \emph{{Extension of $N=2$ Superconformal Algebra and Calabi-yau
  Compactification}}, Mod.Phys.Lett. \textbf{A4} (1989), 557.

\bibitem{Odake:1989ev}
\bysame, \emph{{C = 3-$d$ Conformal Algebra With Extended Supersymmetry}},
  Mod.Phys.Lett. \textbf{A5} (1990), 561.

\bibitem{Polyakov:1970xd}
Alexander~M. Polyakov, \emph{{Conformal symmetry of critical fluctuations}},
  JETP Lett. \textbf{12} (1970), 381--383.

\bibitem{Polyakov:1974gs}
A.M. Polyakov, \emph{{Nonhamiltonian approach to conformal quantum field
  theory}}, Zh.Eksp.Teor.Fiz. \textbf{66} (1974), 23--42.

\bibitem{Qualls:2013eha}
Joshua~D. Qualls and Alfred Shapere, \emph{{Bounds on Operator Dimensions in 2D
  Conformal Field Theories}},  (2013).

\bibitem{Rattazzi:2008pe}
Riccardo Rattazzi, Vyacheslav~S. Rychkov, Erik Tonni, and Alessandro Vichi,
  \emph{{Bounding scalar operator dimensions in 4D CFT}}, JHEP \textbf{0812}
  (2008), 031.

\bibitem{Witten:2007kt}
Edward Witten, \emph{{Three-Dimensional Gravity Revisited}},  (2007).

\bibitem{Yau:2000}
S.-T. Yau, \emph{Review of geometry and analysis}, Asian J. Math. \textbf{4}
  (2000), no.~1, 235--278, Kodaira's issue.

\end{thebibliography}

\end{document}